\documentclass[%
 reprint,
superscriptaddress,
bibnotes,
 amsmath,amssymb,
 aps,
prb,
]{revtex4-1}

\usepackage{graphicx}
\usepackage{dcolumn}
\usepackage{bm}
\usepackage{color}
\definecolor{OliveGreen}{rgb}{0,0.6,0}
\definecolor{auburn}{rgb}{0.43, 0.21, 0.1}
\definecolor{blue_violet}{rgb}{0.54, 0.17, 0.89}

\usepackage{appendix}
\usepackage{chngcntr}
\usepackage{etoolbox}
\usepackage{lipsum}

\AtBeginEnvironment{appendices}{%
\addcontentsline{toc}{section}{Appendices}
\counterwithin{figure}{section}
\counterwithin{equation}{section}
\counterwithin{table}{section}
}

\begin{document}


\title{Kinetic Pathways of Phase Decomposition Using Steepest-Entropy-Ascent Quantum 
Thermodynamics Modeling. Part\;II: Phase Separation and Ordering}

\author{Ryo Yamada}
\email{ryo213@vt.edu}
\affiliation{Materials Science and Engineering Department, Virginia Polytechnic Institute and State University, Blacksburg, Virginia 24061, USA}
\author{Michael R. von Spakovsky}
\email{vonspako@vt.edu}
\affiliation{Center for Energy Systems Research, Mechanical Engineering Department, Virginia Polytechnic Institute and State University, Blacksburg, Virginia 24061, USA}
\author{William T. Reynolds, Jr.}
\email{reynolds@vt.edu}
\affiliation{Materials Science and Engineering Department, Virginia Polytechnic Institute and State University, Blacksburg, Virginia 24061, USA}

\date{\today}

\begin{abstract}
The kinetics of ordering and concurrent ordering and clustering is analyzed with an equation of motion initially developed to account for dissipative processes in quantum systems. A simplified energy eigenstructure, or pseudo-eigenstructure, is constructed from a static concentration wave method to describe the configuration-dependent energy for atomic ordering and clustering in a binary alloy. This pseudo-eigenstructure is used in conjunction with an equation of motion that follows steepest entropy ascent to calculate the kinetic path that leads to ordering and clustering in a series of hypothetical alloys. By adjusting the thermodynamic solution parameters, it is demonstrated that the model can predict the stable equilibrium state as well as the unique thermodynamic path and kinetics of continuous/discontinuous ordering and concurrent processes of simultaneous ordering and phase separation.
\end{abstract}

\pacs{Valid PACS appear here}
\maketitle

\section{\label{chap6_sec:level1}Introduction}
As discussed in Part\;I \cite{yamada2018kineticpartI}, the decomposition of a thermodynamically unstable solution into two stable phases can take place by kinetic pathways differentiated by the spatial distribution of the phases. The equation of motion within the steepest-entropy-ascent quantum thermodynamic (SEAQT) framework provides a means of predicting the operative kinetic pathway between the limiting cases of continuous (spinodal phase separation) and discontinuous (nucleation and growth) transformations.

An additional degree of freedom can be considered during decomposition of solid-solutions. For a binary A--B alloy, the two chemical species can order or cluster on the crystalline lattice irrespective of whether the transformation pathway is continuous or discontinuous. The preference for ordering or clustering is determined by the relative chemical affinities of the solution components. When the two components have a stronger chemical affinity for each other than for themselves, the stable ground-state structure will tend toward the ordering of A and B on the crystal lattice. On the other hand, if the two components prefer to bond to themselves rather than each other, the ground state will be characterized by the clustering of A and B into two chemically distinct phases (i.e., phase separation). 

This general tendency can be quantified by {\it effective} pairwise interaction energies, $w_n=V^{(n)}_{AA}+V^{(n)}_{BB}-2V^{(n)}_{AB}$, where $V^{(n)}_{ij}$ is the component-specific $n^{th}$-nearest-neighbor pair interaction energy ($i,j=$A or B). The effective pairwise interaction energies, $w_n$, are convenient parameters characterizing chemical affinity in a solid. Considering only 1st-nearest-neighbor pair interaction energies, the interaction energy can be defined such that $w_1>0$ produces ordering at the ground state and $w_1<0$ produces clustering.  Although the 1st-nearest-neighbor interactions are the largest contribution to the chemical affinity, more distant interaction energies can be influential. As discussed in references \cite{inden1974ordering,ino1978pairwise,soffa2010interplay}, a competition between ordering and phase separation is expected when the 1st-neighbor and 2nd-neighbor effective interaction energies have opposite signs (i.e., when $w_1>0$ and $w_2<0$ or when $w_1<0$ and $w_2>0$). Experimentally, such concurrent phase separation and ordering is well documented in Fe--Be \cite{ino1978pairwise}, Al--Li, and Ni--Al alloys \cite{soffa1989decomposition}. However, current theoretical frameworks to model the kinetics of concurrent transformations come with significant limitations.

For example, since molecular dynamics simulations \cite{oramus2003ordering} are based on classical mechanics, their use cannot be justified below the Debye temperature, and they are limited to a time scale that is extremely short relative to diffusional processes. Although kinetic Monte Carlo methods \cite{kessler2003ordering} and phase field models \cite{ichitsubo2000kinetics,proville2001kinetics} can simulate much longer times, they are, respectively, based on stochastic and phenomenological thermodynamics. Physical insights tend to be lost with stochastic methods, and phenomenological approaches are not strictly applicable far from equilibrium because they utilize a local/near equilibrium assumption.  

The Path Probability Method (PPM) \cite{kikuchi1966path} --- an extension of the Cluster Variation Method (CVM) \cite{kikuchi1951theory} to time domains --- can describe kinetic paths from an initial non-equilibrium state to an equilibrium state without relying on a stochastic approach or a local/near equilibrium assumption. The time-evolution of state is taken to be the most probable kinetic path determined by maximizing a path probability function defined for all possible paths of the transformation process. It is known that states derived for the long-time limit in PPM converge to the equilibrium  predicted by CVM, and that the calculated kinetic paths significantly deviate from the steepest descent direction of the free-energy contour surface \cite{mohri1996kinetic}. However, the PPM calculation has some drawbacks. Many path variables make it computationally demanding, alloy composition is not automatically conserved during the kinetic calculations, and it has not been extended beyond applications involving single-phase alloys \cite{mohri1996kinetic}. 

These aforementioned limitations can be circumvented with the SEAQT framework developed in Part\;I \cite{yamada2018kineticpartI}. With this approach, a unique kinetic path is simply determined from an arbitrary initial state to stable equilibrium by an equation of motion that follows the direction of steepest entropy ascent with an alloy composition fixed. 
While continuous and discontinuous transformations are investigated in Part\;I, concurrent phase separation and ordering is explored here using the SEAQT theoretical framework. The paper is organized as follows. In Sec.\;\ref{chap6_sec:level2}, a simplified energy eigenstructure, or pseudo-eigenstructure, is constructed for a lattice that can undergo ordering or clustering using the static concentration wave method \cite{khachaturyan2013theory}. In Sec.\;\ref{chap6_sec:level3}, hypothetical alloys that are expected to undergo different decomposition mechanisms are constructed by adjusting the relative values of pair interaction energies, and time-evolution processes are calculated for each alloy system. Finally, the predicted kinetic pathways from SEAQT are summarized in Sec.\;\ref{chap6_sec:level4}.

\section{\label{chap6_sec:level2}Theory}
\begin{figure}
\begin{center}
\includegraphics[scale=0.40]{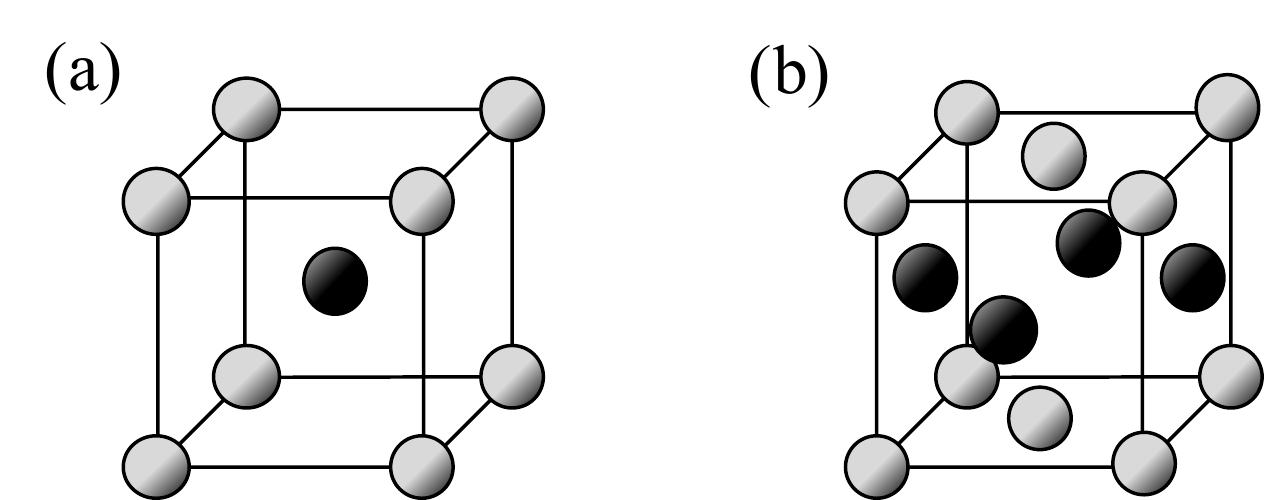}
\caption{\label{fig6:B2_L10} The (a) B2 and (b) L1$_0$ lattices. While the B2 structure has a body-centered cubic (BCC) structure, the L1$_0$ structure has a face-centered cubic (FCC) structure. The gray and black particles represent different atomic species. }
\end{center}
\end{figure}

The SEAQT equation of motion and procedure for specifying initial states are the same as described in Part\;I \cite{yamada2018kineticpartI}. Whereas the pseudo-eigenstructure constructed from a reduced-order method (i.e., mean-field approximation) for the alloy system in Part\;I is parameterized by the concentration of B-type atoms,  ordering and clustering behavior require a description of the system that includes both concentration and the long-range order (LRO) parameter (see the schematic of Fig.\;1 in Part\;I). The pseudo-eigenstructure is constructed here by employing the static concentration wave (SCW) method \cite{khachaturyan2013theory}, which is a type of mean-field approximation. As an illustration, a pseudo-eigenstructure for systems that have a B2 or L1$_0$ lattice (Fig.\;\ref{fig6:B2_L10}) at low temperatures is constructed in Sec.\;\ref{chap6_sec:level2_1}. In addition, an expected phase transformation behavior depending on thermodynamic solution parameters at the ground state is given in Sec.\;\ref{chap6_sec:level2_2}.

\subsection{\label{chap6_sec:level2_1}Pseudo-eigenstructure}
As described in Part\;I \cite{yamada2018kineticpartI}, the configurational energy in a binary alloy system is given by \cite{khachaturyan2013theory}
\begin{equation}
E=\frac{1}{2} \sum_{\bm{\mathrm{r}},\bm{\mathrm{r}}'} w(\bm{\mathrm{r}} - \bm{\mathrm{r}}') n(\bm{\mathrm{r}}) n(\bm{\mathrm{r}}') \; ,   \label{eq6:total_energy_MF_original}
\end{equation}
where $w(\bm{\mathrm{r}} - \bm{\mathrm{r}}')$ is a pairwise interatomic interaction energy between two atoms at lattice sites $\bm{\mathrm{r}}$ and $\bm{\mathrm{r}}'$ and $n(\bm{\mathrm{r}})$ and $n(\bm{\mathrm{r}}')$ are the distribution functions at those lattice points. In the SCW method, the distribution functions, $n(\bm{\mathrm{r}})$, for the B2 and L1$_0$ lattices are given as \cite{khachaturyan2013theory}
\begin{equation}
n(\bm{\mathrm{r}})=c+\frac{1}{2}\eta e^{i\bm{\mathrm{k}}_0 \cdot \bm{\mathrm{r}}}  ,    \label{eq6:occupation_probability_static_conc}
\end{equation}
where $c$ is the concentration of B-type atoms and $\bm{\mathrm{k}}_0$ are the wave vectors of special-points, which correspond to the (111) and (001) points for the B2 and L1$_0$ structures, respectively. From Eqs.\;(\ref{eq6:total_energy_MF_original}) and (\ref{eq6:occupation_probability_static_conc}), the configurational energy for either lattice becomes \cite{khachaturyan2013theory,cheong1994thermodynamic}
\begin{equation}
E(c,\eta)=N\left[ \frac{1}{2} c (1-c) V(\bm{0}) - \frac{1}{8} \eta^2 V(\bm{\mathrm{k}}_0)  \right],   \label{eq6:total_energy_static_conc}
\end{equation}
where $N$ is the number of atoms in a system, and $V(\bm{0})$ and $V(\bm{\mathrm{k}}_0)$ are, respectively, given by
\begin{equation}
\begin{split}
V^{\mathrm{BCC}}(\bm{0})=8w_1+6w_2+12w_3+24w_4+\cdot \cdot \cdot \quad  \\
V^{\mathrm{FCC}}(\bm{0})=12w_1+6w_2+24w_3+12w_4+\cdot \cdot \cdot \quad    \label{eq6:interaction_energies_static_conc_V0}
\end{split} 
\end{equation}
and
\begin{equation}
\begin{split}
V^{\mathrm{B2}}(\bm{\mathrm{k}}_0)=-8w_1+6w_2+12w_3-24w_4+\cdot \cdot \cdot  \quad \\
V^{\mathrm{L1}_0}(\bm{\mathrm{k}}_0)=-4w_1+6w_2-8w_3+12w_4+\cdot \cdot \cdot  \quad \; .  \label{eq6:interaction_energies_static_conc_V1}
\end{split} 
\end{equation}
Here the $w_{n}$ represent the $n^{th}$-nearest-neighbor effective pair interaction energies. 

The energy of the L1$_0$ lattice has an additional complication in that it can involve a tetragonal distortion. When this distortion is taken into account, the energy shown in Eq.\;(\ref{eq6:total_energy_static_conc}) becomes \cite{cheong1994thermodynamic}
\begin{equation}
E(c,\eta)=N\left[ \frac{1}{2} c (1-c) V(\bm{0}) - \frac{1}{8} \eta^2 V(\bm{\mathrm{k}}_0) - \eta^4 e \right],   \label{eq6:total_energy_static_conc_L10}
\end{equation}
where $e$ is the contribution of an elastic strain energy stemming from a tetragonal distortion given by
\begin{equation}
e=v C_{11} \{ (1+\frac{C_{12}}{C_{11}}) (\bar{\epsilon}^0_{11})^2 +2 \frac{C_{12}}{C_{11}} \bar{\epsilon}^0_{11} \bar{\epsilon}^0_{33} + \frac{1}{2} (\bar{\epsilon}^0_{33})^2 \}  ,
\label{eq6:tetragonal_distortion_energy_static_conc}
\end{equation}
where $v$ is the average atomic volume, $C_{ij}$ are the elastic constants, and $\bar{\epsilon}^0_{ij}$ are the average strains.

The degeneracy of the energy in Eq.\;(\ref{eq6:total_energy_static_conc}) (or Eq.\;(\ref{eq6:total_energy_static_conc_L10})) is given by a binomial coefficient as
\begin{equation}
\begin{split}
g(c,\eta) =\frac{(N^{\alpha}) !}{(N^{\alpha}_{\mbox{\scriptsize A}})! \cdot (N^{\alpha}_{\mbox{\scriptsize B}})!} \cdot \frac{(N^{\beta}) !}{(N^{\beta}_{\mbox{\scriptsize A}})! \cdot (N^{\beta}_{\mbox{\scriptsize B}}!)}  \\
=\frac{ \left( \frac{N}{2} \right) !}{(N^{\alpha}_{\mbox{\scriptsize A}} )! \cdot ( \frac{N}{2} - N^{\alpha}_{\mbox{\scriptsize A}}) !} \cdot \frac{ \left( \frac{N}{2} \right) !}{ ( N^{\beta}_{\mbox{\scriptsize A}}) ! \cdot (\frac{N}{2} - N^{\beta}_{\mbox{\scriptsize A}} )!}  ,     \label{eq6:degeneracy_SCW}
\end{split} 
\end{equation}
where $\alpha$ and $\beta$ indicate the sublattice on which A and B atoms predominate in the ordered lattice, and $N^{\alpha}_{\mbox{\scriptsize A}}$ and $N^{\beta}_{\mbox{\scriptsize A}}$ are, respectively, given by
\begin{equation}
\begin{split}
N^{\alpha}_{\mbox{\scriptsize A}}=N \left[ \frac{1}{2} (1-c) +\frac{1}{4} \eta \right] , \\
N^{\beta}_{\mbox{\scriptsize A}}=N \left[ \frac{1}{2} (1-c) -\frac{1}{4} \eta \right] .     \label{eq6:degeneracy_SCW_part}
\end{split} 
\end{equation}

The energy represented by Eq.\;(\ref{eq6:total_energy_static_conc}) (or Eq.\;(\ref{eq6:total_energy_static_conc_L10})) forms a continuous or infinite spectrum of energies or energy eigenlevels for the system from which a pseudo-eigenstructure of finite discrete levels can be constructed using the density of states method developed by  Li and von Spakovsky\cite{li2016steepest} (see Part\;I). This set is used by the SEAQT equation of motion to accurately predict the kinetics of system state evolution. The energy eigenlevels, degeneracies, concentrations of B-type atoms, and LRO parameters for the system are then given by
\begin{equation}
E_{i,j} = \frac{1}{g_{i,j}} \int_{\bar{\eta}_{j}}^{\bar{\eta}_{j+1}} \int_{\bar{c}_{i}}^{\bar{c}_{i+1}} g(c, \eta) E (c,\eta) \; dc \; d\eta   \;,  \label{eq6:energy_eigenvalue_pseudo}
\end{equation}
\begin{equation}
g_{i,j}=\int_{\bar{\eta}_{j}}^{\bar{\eta}_{j+1}} \int_{\bar{c}_{i}}^{\bar{c}_{i+1}} g(c, \eta) \; dc \; d\eta \;,   \label{eq6:degeneracy_pseudo}
\end{equation}
\begin{equation}
c_{i,j} = \frac{1}{g_{i,j}} \int_{\bar{\eta}_{j}}^{\bar{\eta}_{j+1}} \int_{\bar{c}_{i}}^{\bar{c}_{i+1}} g(c,\eta) c \; dc \; d\eta \;,    \label{eq6:fraction_down_spin_pseudo}
\end{equation}
and
\begin{equation}
\eta_{i,j}= \frac{1}{g_{i,j}} \int_{\bar{\eta}_{j}}^{\bar{\eta}_{j+1}} \int_{\bar{c}_{i}}^{\bar{c}_{i+1}} g(c,\eta) \eta \; dc \; d\eta \; .    \label{eq6:fraction_down_spin_pseudo}
\end{equation}
The $\bar{c}_{i}$ and $\bar{\eta}_{j}$ are prepared as
\begin{equation}
\bar{c}_{i}= \frac{i}{R_c} \;\;\; \mathrm{and} \;\;\;  \bar{\eta}_{j}= \frac{j}{R_{\eta}} \; ,\label{eq6:concentraion_eigenvalues_and_long-range-parameters}
\end{equation}
where $R_c$ and $R_{\eta}$ are, respectively, the number of intervals in the pseudo-eienstructure for the concentration of B atoms and the LRO parameter, and $i$ and $j$ are integer values ($i=0,1,2,...R_c$ and $j=0,1,2,...R_{\eta}$). The number of intervals, $R_c$ and $R_{\eta}$, is determined by ensuring the following condition is satisfied \cite{yamada2018steepest}:
\begin{equation}
\frac{1}{\beta}\gg \frac{| E_{k+1}-E_k | }{N}  \; ,   \label{eq6:quasi_continuous_condition}
\end{equation}
where the subscripts for each energy eigenlevel is expressed by $k$ (i.e., $E_{i,j} \rightarrow E_k$). 
Since there is a maximum value of accessible LRO parameters for each concentration of B atoms (i.e., $\eta_{\mbox{\scriptsize max}}=2c$ and $2(1-c)$ for $c \leq 0.5$ and $c>0.5$, respectively), the inaccessible LRO parameters are eliminated from the pseudo-eigenstructure. 

Note that the B2 and L1$_0$ lattices are described by the ordering of an underlying disordered lattice (body-centered or face-centered cubic) through the use of a single LRO parameter, $\eta$, in the SCW approach. Pseudo-eigenstructures for alloy systems with different ordered lattices described by more than two LRO parameters can be derived by following a similar procedure. 

The approach used here to describe ordering is equivalent to the Bragg-Williams approximation or the point-approximation \cite{kikuchi1997,mohri2013cluster} of the cluster variation method. More elaborate mean-field approximations for atomic configurations are available in the cluster variation method \cite{kikuchi1951theory,kikuchi1997,mohri2013cluster} that incorporate short-range correlations through the use of defined cluster configurations, but the point-approximation is employed here for simplicity.

\subsection{\label{chap6_sec:level2_2}Ground-state analysis}
In Sec.\;\ref{chap6_sec:level2_1}, pseudo-eigenstructures are constructed for alloys exhibiting one of two types of ordering: L1$_0$ ordering from an initially disordered FCC solid-solution (or FCC$_{s.s.}$) and B2 ordering from an initially disordered BCC solid-solution (or BCC$_{s.s.}$). When $V(\bm{0})$, $V(\bm{\mathrm{k}}_0)$, and $e$ are assumed to be constant, the energy, Eq.\;(\ref{eq6:total_energy_static_conc_L10}), can be written, using the reduced parameters $\omega$ and $\alpha$, as
\begin{equation}
\frac{E(c,\eta)}{V(\bm{\mathrm{k}}_0)}=N\left[ \frac{1}{2} c (1-c) \omega - \frac{1}{8} \eta^2 - \eta^4 \alpha \right] ,    \label{eq6:total_energy_static_conc_2}
\end{equation}
where
\begin{equation}
\omega \equiv \frac{V(\bm{0})}{V(\bm{\mathrm{k}}_0)} , \;\;\;\;\;\;\;  \alpha \equiv \frac{e}{V(\bm{\mathrm{k}}_0)}  ,\label{eq6:reduce_parameter_static_conc}
\end{equation}
and it is assumed that $V(\bm{\mathrm{k}}_0)>0$. The parameter $\alpha$ describes the contribution of strain energy. When $\alpha=0$, there is no tetragonal distortion and no contribution to the strain energy (which is the case with B2 ordering). The parameter $\omega$ reflects the relative contributions of the $n^{th}$-nearest-neighbor (effective) interaction energies (see Eqs.\;(\ref{eq6:interaction_energies_static_conc_V0}) and (\ref{eq6:interaction_energies_static_conc_V1})). 

\begin{figure}
\begin{center}
\includegraphics[scale=0.5]{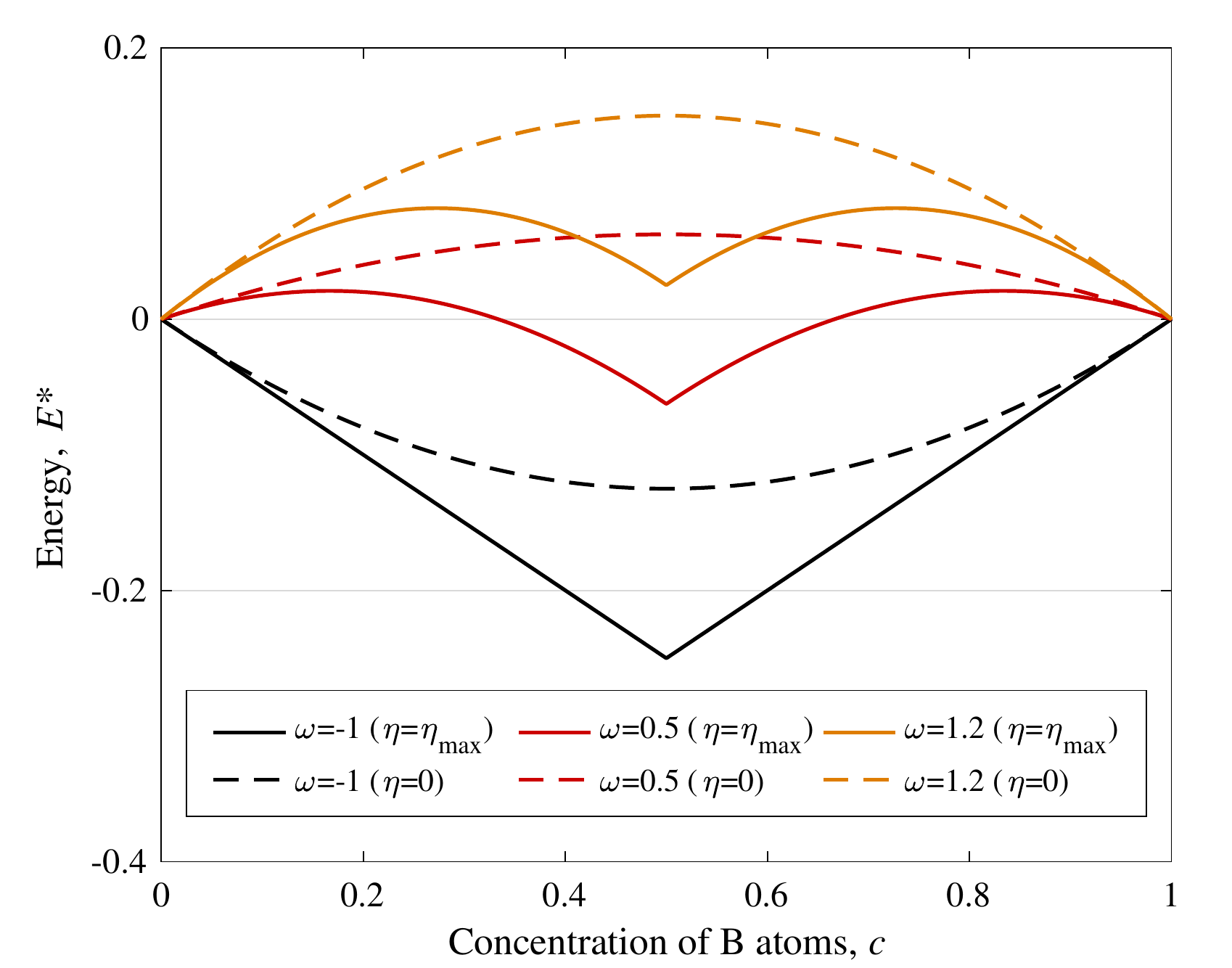}
\caption{\label{fig6:ground_state_analysis} The ground-state analysis for three representative alloy systems ($\omega=-1.0, \;0.5, \; \mathrm{and} \; 1.2$) with $\alpha=0$, which correspond to Alloys 3, 4, and 5, respectively, in the subsequent calculations. The color represents the different alloys. While the energies of the corresponding solid-solutions ($\eta=0$) are shown by the dashed curves, fully ordered phases ($\eta=\eta_{\mathrm{max}}$) are shown by the solid lines. Here, the energies are normalized as $E^* = E(c,\eta)/V(\bm{\mathrm{k}}_0)$.}
\end{center}
\end{figure}

Three representative relations between energy and concentration of B atoms for alloy systems, $\omega=-1.0, \;0.5, \; \mathrm{and} \; 1.2$, with $\alpha=0$ are shown in Fig.\;\ref{fig6:ground_state_analysis}. In the figure, $E^* = E(c,\eta)/V(\bm{\mathrm{k}}_0)=0$ corresponds to the segregation limit --- a straight line connecting the energies of phases composed of pure A atoms and pure B atoms. The segregation limit shows the ground-state energy of a phase-separated configuration and facilitates comparisons with solid-solutions and ordered phases. The broken lines in the figure are energies of solid-solutions ($\eta=0$); if the energy of a solid-solution at a given concentration is above (or below) the segregation limit, the system prefers phase separation (or solid-solution). The energies of a solid-solution are decreased by the ordering term in Eq.\;(\ref{eq6:total_energy_static_conc_2}), and the energies of a fully ordered phase ($\eta=\eta_{\mathrm{max}}$) are shown by the solid curves in the figure. Relative to the segregation limit, systems with $\omega<0$ (black lines) will always tend to order, but systems with $\omega>0$ (red and orange curves) {\it may} order, or cluster, or do both. The reduced parameters, $\omega$ and $\alpha$, are used to represent hypothetical alloy systems with these different behaviors in the following calculations. 

\section{\label{chap6_sec:level3}Results and Discussion}
Phase decomposition in a binary system whose chemical affinity leads to phase separation (a single solid-solution decomposing into two different solid-solutions) is discussed in Part\;I \cite{yamada2018kineticpartI}. Additional types of decomposition can be produced by adjusting pairwise interaction energies to simulate hypothetical alloy systems that favor ordering (Sec.\;\ref{chap6_sec:level3_1}) or both phase separation and ordering (Sec.\;\ref{chap6_sec:level3_2}). The kinetic pathways in these hypothetical alloys from some initial unstable phase to stable equilibrium phases are explored using the SEAQT model. All the time scales in the subsequent calculations are normalized by a relaxation time, $\tau$, but they can be correlated to a real time by following a similar procedure to that shown in Part\;I.

\subsection{\label{chap6_sec:level3_1}Ordering}
The values of $\omega$ and $\alpha$ for three hypothetical alloys that exhibit ordering are shown in Table\;\ref{table6:ordering_three_cases}. These correspond to the values used in reference \cite{cheong1994thermodynamic}. Alloy-1 and Alloy-2 are for alloys that involve L1$_0$ ordering on an FCC lattice and Alloy-3 represents B2 ordering on a BCC lattice (because there can be no tetragonal distortion in this case, $\alpha=0$). Phase diagrams calculated using these parameters in the SEAQT theoretical framework (see Appendix\;\ref{chap6_sec:level5_1}) are shown in Fig.\;\ref{fig6:phase_diagram_ordering_three_cases}. For all the kinetic calculations, the initial temperature was chosen to be $T^{*}_0=0.5$ (normalized temperature $T^*=k_B T/V(\bm{\mathrm{k}}_0$)) with fluctuations in the initial state generated using $N_0=10^2$ (see Part\;I \cite{yamada2018kineticpartI} for details). 
\begin{table}
\begin{center}
\caption{\label{table6:ordering_three_cases} The assumed values of $\omega$ and $\alpha$ in the three model alloy systems for the ordering calculations in Sec.\;\ref{chap6_sec:level3_1}.  Alloy-1 and Alloy-2 are alloy systems involving L1$_0$ ordering on an FCC lattice and Alloy-3 represents B2 ordering on a BCC lattice. These values correspond to those used in reference \cite{cheong1994thermodynamic}.}
\begin{tabular}{ c  c  c } 
$\quad$ & $\quad$ & $\quad$ \\   \hline \hline
$\quad\quad\quad$  & \quad $\omega$ \quad \quad & \quad $\alpha$ \quad \quad \\ \hline
\quad Alloy-1 (L1$_0$ on FCC) \quad & -1.0 & 0.10 \\
\quad Alloy-2 (L1$_0$ on FCC) \quad & -1.0 & 0.05 \\ 
\quad Alloy-3 (B2 on BCC) \quad & -1.0 & 0.00 \\ \hline \hline
\end{tabular}
\end{center}
\end{table}
\begin{figure}
\begin{center}
\includegraphics[scale=0.28]{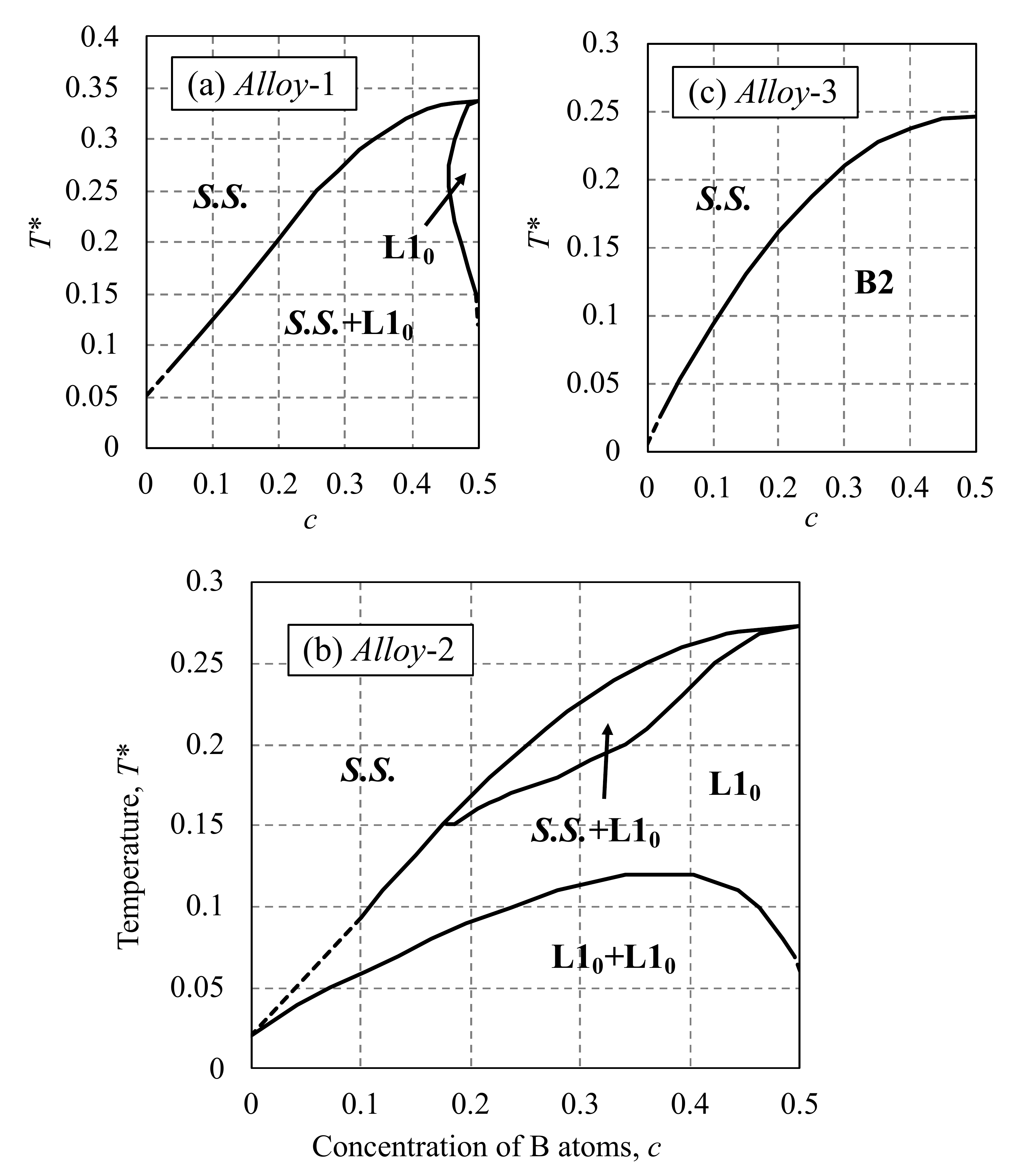}
\caption{\label{fig6:phase_diagram_ordering_three_cases} The phase diagrams of the three model alloy systems in Sec.\;\ref{chap6_sec:level3_1} calculated using the SEAQT model. The diagrams qualitatively agree with the ones determined by a free-energy analysis \cite{cheong1994thermodynamic}. While there are two-phase regions, FCC$_{s.s.}$\;$+$\;L1$_0$, in Alloy-1 and Alloy-2, there are only single phase regions, BCC$_{s.s.}$ or B2, in Alloy-3. Here the temperatures are normalized as $T^*=k_BT/V(\bm{\mathrm{k}}_0)$, and the estimated lines are shown as broken lines. }
\end{center}
\end{figure}

\subsubsection{\label{chap6_sec:level3_1-1}Alloy-1 (FCC$_{s.s.}$ $\Rightarrow$ L1$_0$)}
\begin{figure}
\begin{center}
\includegraphics[scale=0.17]{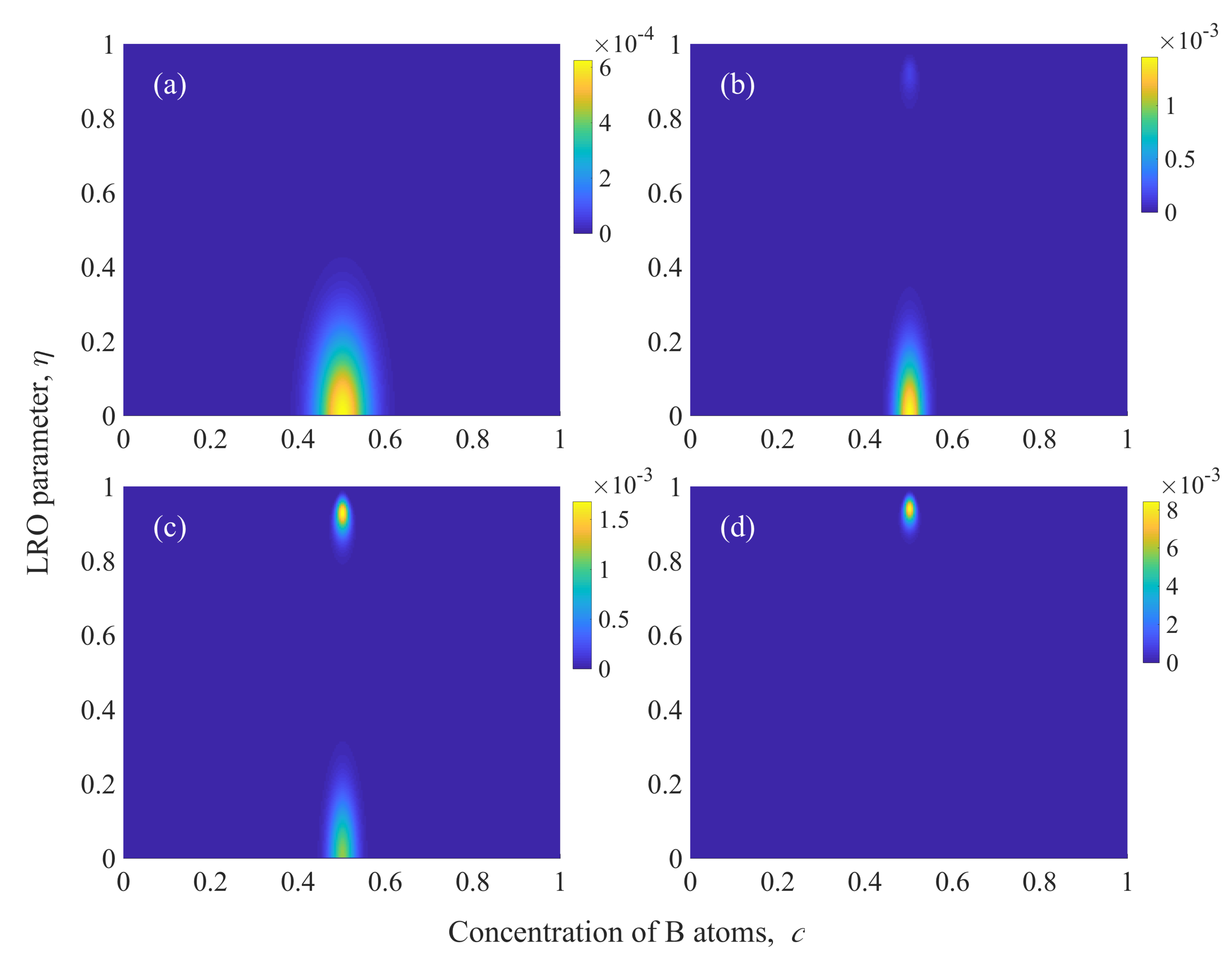}
\caption{\label{fig6:kinetic_ordering_case_B_1_1} The calculated kinetic ordering process in a A--50.0\;at.\%\;B alloy (Alloy-1 in Sec.\;\ref{chap6_sec:level3_1}) at $T^{*}_R=0.30$ using $N=10^4$ and $N_0=100$. Each panel represents a snapshot corresponding to a normalized time, $t^*$, of (a) $0.0000$, (b) $0.0220$, (c) $0.0260$, and (d) $0.0350$.  }
\end{center}
\end{figure}

\begin{figure}
\begin{center}
\includegraphics[scale=0.17]{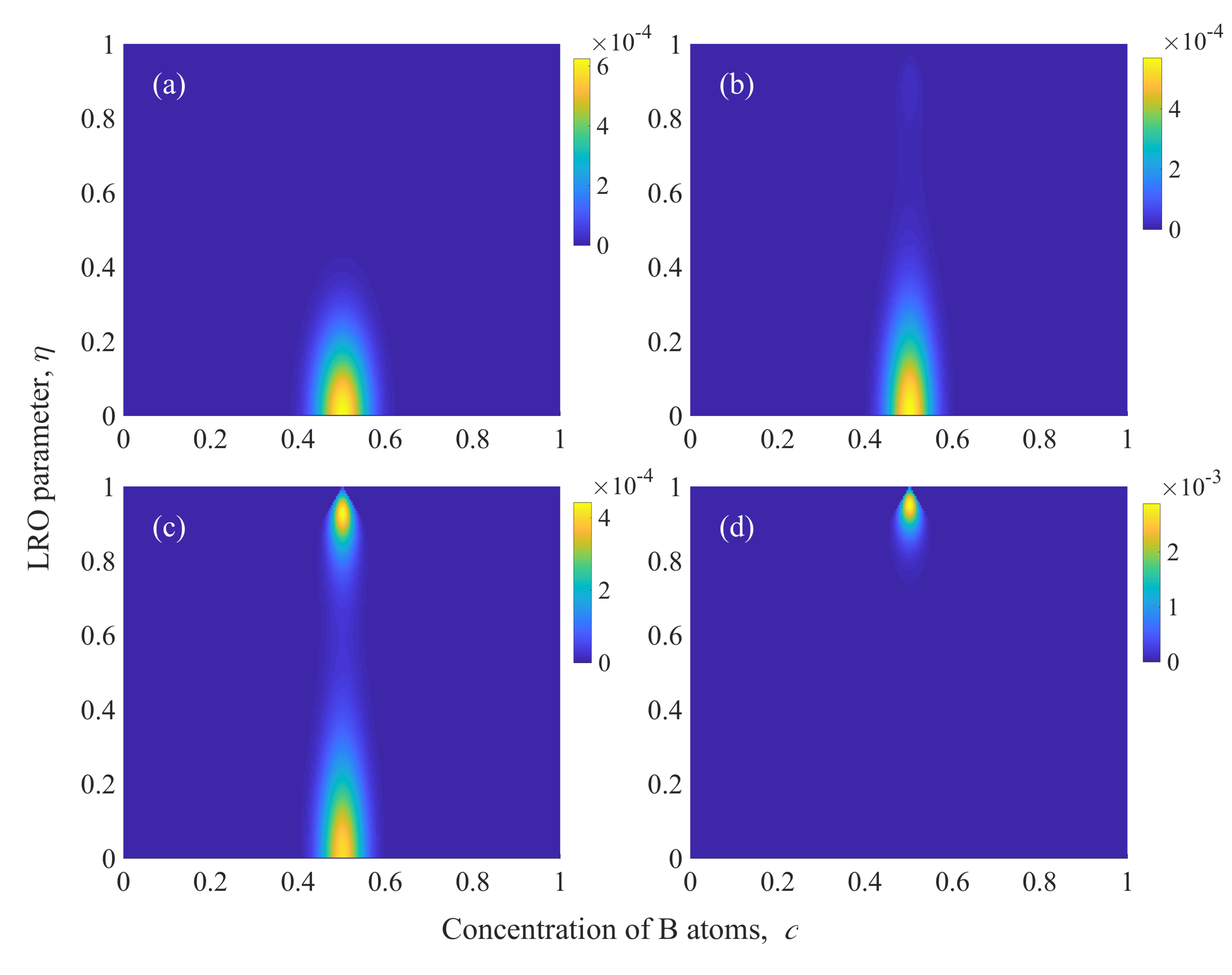}
\caption{\label{fig6:kinetic_ordering_case_B_1_2} The calculated kinetic ordering process in a A--50.0\;at.\%\;B alloy (Alloy-1 in Sec.\;\ref{chap6_sec:level3_1}) at $T^{*}_R=0.15$ using $N=10^4$ and $N_0=100$. Each panel represents a snapshot corresponding to a normalized time, $t^*$, of (a) $0.0000$, (b) $0.0021$, (c) $0.0026$, and (d) $0.0032$. }
\end{center}
\end{figure}

The calculated kinetic ordering processes from a single solid-solution in a A--50.0\;at.\%\;B alloy at two different annealing temperatures, $T^{*}_R=0.30$ and $T^{*}_R=0.15$, are, respectively, shown in Figs.\;\ref{fig6:kinetic_ordering_case_B_1_1} and \ref{fig6:kinetic_ordering_case_B_1_2} for the ordering system. Each panel in these figures represents a state of the system at a particular instant of time.  The concentration of the B-type atom in the alloy, $c$, varies along the horizontal axis, and the long-range order parameter, $\eta$, varies along the vertical axis. The color of each pixel in a panel represents the probability of the combination of concentration and LRO corresponding to the pixel's location. The sum of the probabilities over all possible configurations of concentration and LRO in a given panel is unity. The change in the configuration probabilities from panel to panel thus shows how the alloy evolves from a chosen initial system state to the equilibrium state. 

The initial states (the upper left panels in Figs.\;\ref{fig6:kinetic_ordering_case_B_1_1} and \ref{fig6:kinetic_ordering_case_B_1_2}) are the same. Ordering in Fig.\;\ref{fig6:kinetic_ordering_case_B_1_1} takes place at a higher temperature (lower driving force) than in Fig.\;\ref{fig6:kinetic_ordering_case_B_1_2}. At the higher temperature of Fig.\;\ref{fig6:kinetic_ordering_case_B_1_1}, the probability distribution shifts discontinuously to the final state. The initial probability distribution (near $c=0.5, \eta=0)$ decreases as the ordered phase suddenly appears (nucleates) near the final state $(c=0.5, \eta=1)$, and order parameters in the intervening range, say $0.5<\eta<0.8$, are essentially zero for all times. This characteristic is a signature of a discontinuous transformation.

At the lower annealing temperature of Fig.\;\ref{fig6:kinetic_ordering_case_B_1_2}, the probability distribution ends up at a similar final state, but it gradually shifts vertically from the initial state corresponding to the disordered solid-solution to the final equilibrium state corresponding to $\eta=1$ in the manner of a continuous transformation. The system traverses all values of the order parameter between $\eta=0$ and $\eta=1$ during the transformation.

\begin{figure}
\begin{center}
\includegraphics[scale=0.21]{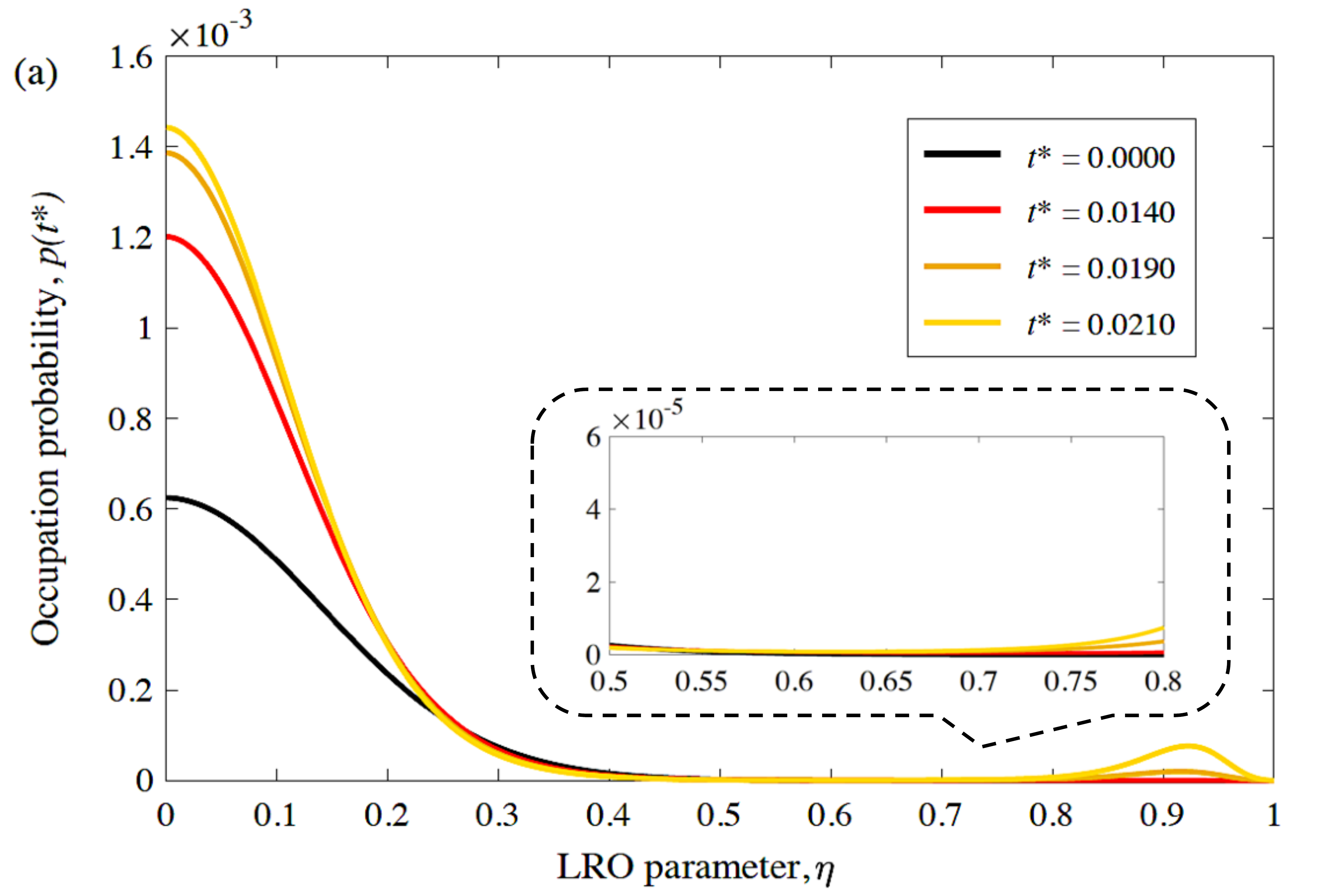}
\includegraphics[scale=0.21]{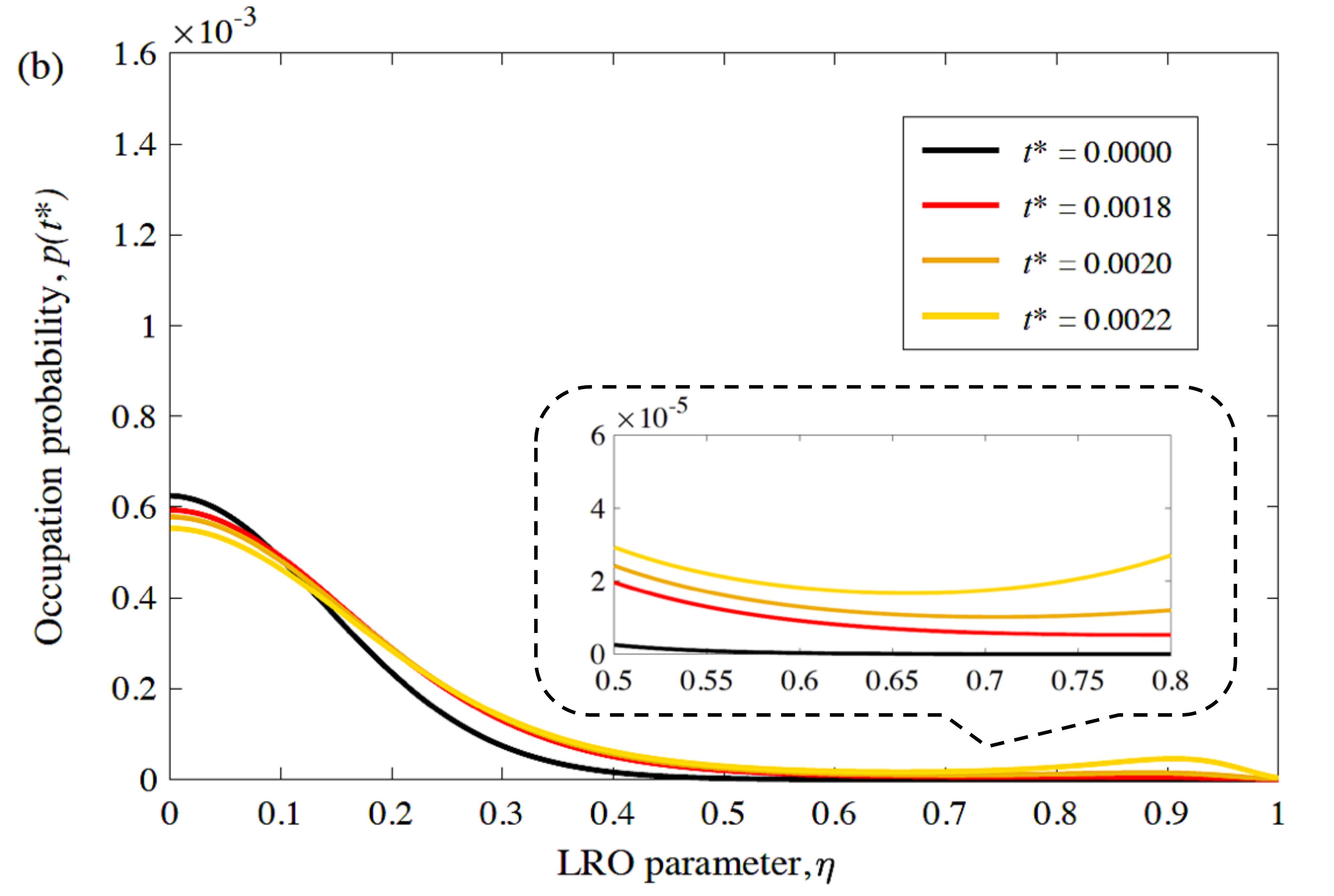}
\caption{\label{fig6:kinetic_ordering_case_B_1_3} The time-dependences of the probability distributions in terms of the LRO parameter at $c=0.5$ in the calculated results shown in Figs.\;\ref{fig6:kinetic_ordering_case_B_1_1} and \ref{fig6:kinetic_ordering_case_B_1_2}; (a) $T^{*}_R=0.30$ and (b) $T^{*}_R=0.15$. }
\end{center}
\end{figure}

The LRO parameter occupancy probabilities for these two cases are shown in Fig.\;\ref{fig6:kinetic_ordering_case_B_1_3} for the same time sequences. Fig.\;\ref{fig6:kinetic_ordering_case_B_1_3}\;(a) corresponds to discontinuous ordering at the higher annealing temperature. Only disordered and nearly fully ordered states are occupied at any time. In Fig.\;\ref{fig6:kinetic_ordering_case_B_1_3}\;(b), which corresponds to continuous ordering, the order parameter gradually traverses all possible ordered states. No nucleation of the stable ordered phase is required. The appearance of different kinetic behavior (continuous versus discontinuous) at different annealing temperatures is also obtained by varying alloy composition (e.g., A--30.0\;at.\%\;B and A--40.0\;at.\%\;B) and by modifying the initial fluctuations as discussed in Part\;I \cite{yamada2018kineticpartI}.

The discontinuous and continuous transformations discussed in Part\;I \cite{yamada2018kineticpartI} are also sometimes called 1st-order and 2nd-order transitions on the Ehrenfest scheme based upon the appearance of a discontinuity in a derivative of the free-energy with respect to a thermodynamic variable like temperature. For 1st-order transitions (discontinuous/nucleation-growth), the free-energy changes abruptly from one phase to another, whereas the free-energy changes smoothly between phases during a 2nd-order (continuous/spinodal) transition. In the context of ordering, both 1st- and 2nd-order transitions have been theoretically and experimentally confirmed for L1$_0$ ordering \cite{tanaka1994spinodal,cheong1994thermodynamic,van1973order,kikuchi1974superposition}. When an annealing temperature is relatively low/high, the transformation shows 2nd-order/1st-order behavior. The continuous behavior seen in Fig.\;\ref{fig6:kinetic_ordering_case_B_1_2} corresponds to a 2nd-order transition (which is also sometimes called spinodal ordering).

\subsubsection{\label{chap6_sec:level3_1_2}Alloy-2 (FCC$_{s.s.}$ $\Rightarrow$ L1$_0$\;+\;L1$_0$)}
The phase diagram in this alloy system, Fig.\;\ref{fig6:phase_diagram_ordering_three_cases}\;(b), shows an interesting phenomenon at low temperatures: a single-ordered phase decomposes into two different ordered phases, each of which has a different composition.

\begin{figure}
\begin{center}
\includegraphics[scale=0.18]{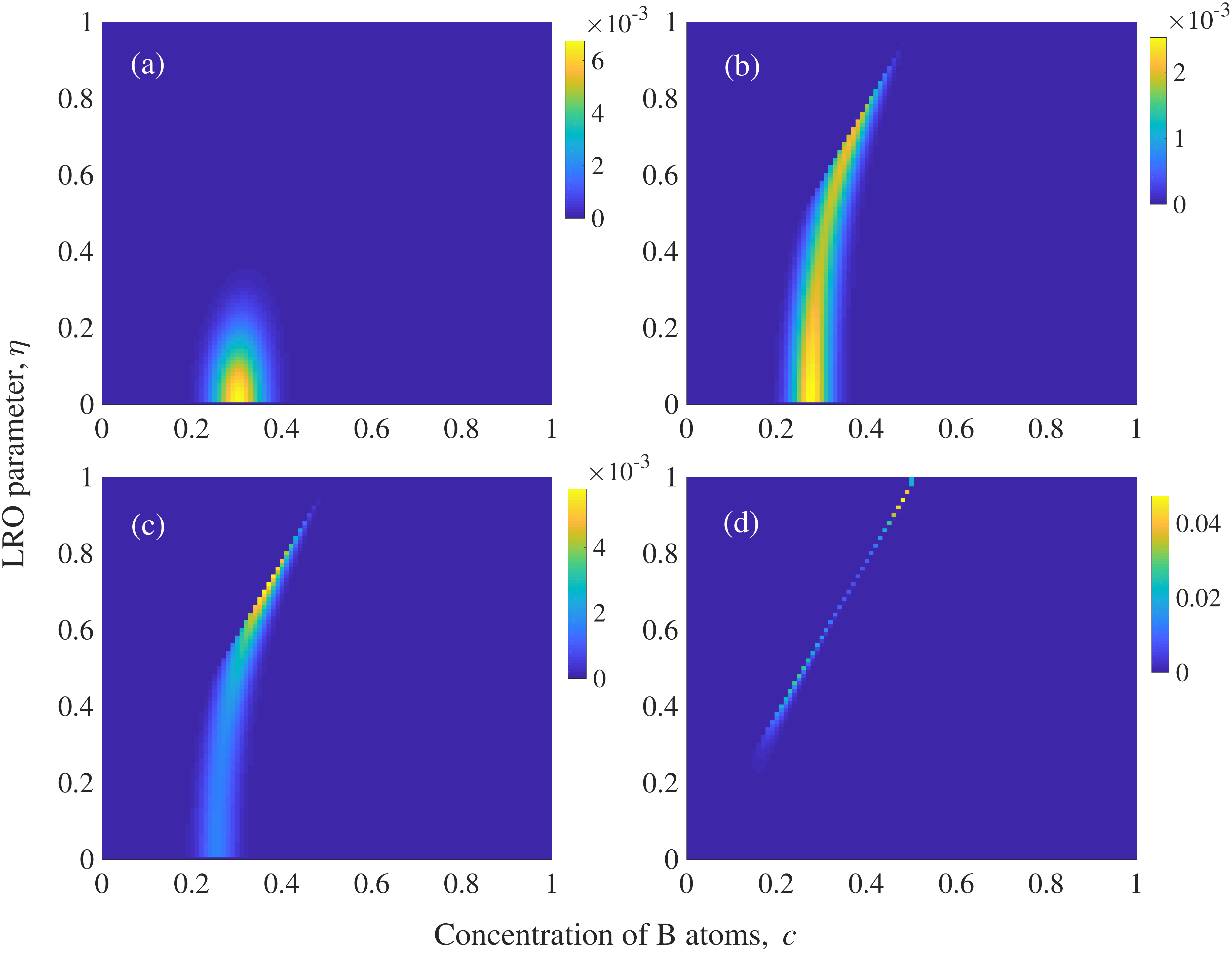}
\caption{\label{fig6:kinetic_ordering_case_B_2} The calculated kinetic evolution process in a A--30.0\;at.\%\;B alloy (Alloy-2 in Sec.\;\ref{chap6_sec:level3_1}) at $T^{*}_R=0.05$ using $N=10^4$ and $N_0=100$. The normalized time, $t^*$, of each snapshot is (a) $0.0000$, (b) $0.0018$, (c) $0.0022$, and (d) $0.0080$, respectively. }
\end{center}
\end{figure}

\begin{figure}
\begin{center}
\includegraphics[scale=0.45]{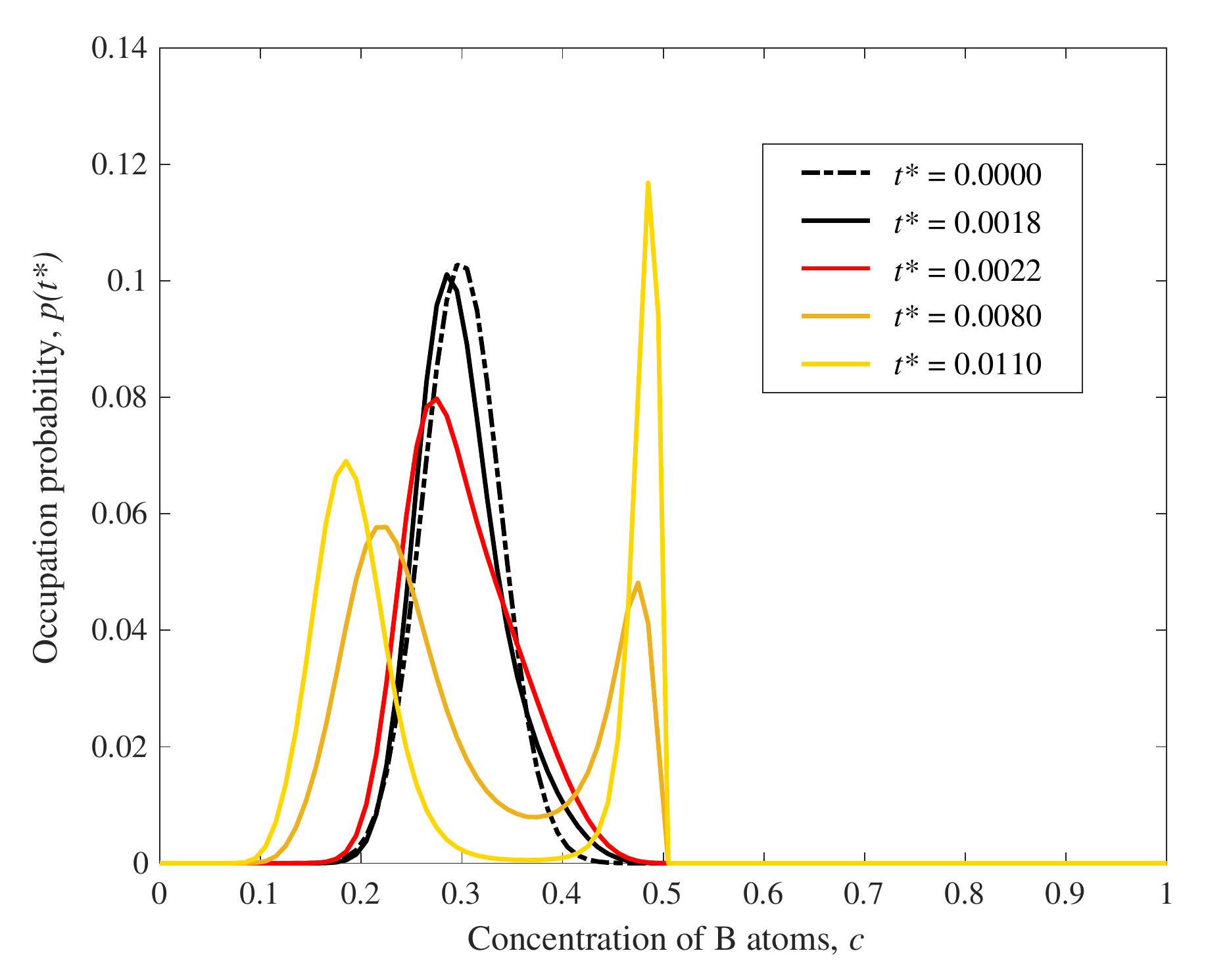}
\caption{\label{fig6:kinetic_ordering_case_B_2_probability} The time-dependence of the probability distributions in terms of the concentration of B-type atoms in the A--30.0\;at.\%\;B alloy at $T^{*}_R=0.05$, which corresponds to the results shown in Fig.\;\ref{fig6:kinetic_ordering_case_B_2}. }
\end{center}
\end{figure}

The calculated kinetic evolution process in a A--30.0\;at.\%\;B alloy at $T^{*}_R=0.05$ is shown in Fig.\;\ref{fig6:kinetic_ordering_case_B_2}. The interesting state parameter in this case is the concentration of B-type atoms. The probability distributions for the concentrations at each instant of time in Fig.\;\ref{fig6:kinetic_ordering_case_B_2} are shown in Fig.\;\ref{fig6:kinetic_ordering_case_B_2_probability}. By the time $t^*=0.0022$ (Fig.\;\ref{fig6:kinetic_ordering_case_B_2}\;(c)), the initial solid-solution has undergone continuous ordering. As time proceeds, the ordered phase decomposes into two different ordered phases with different compositions, Fig.\;\ref{fig6:kinetic_ordering_case_B_2}\;(d). Thus, the kinetic ordering pathway could be described as ``a solid-solution $\Rightarrow$ an ordered phase $\Rightarrow$ two ordered phases". This sequence is the result of the kinetic evolution predicted by the SEAQT equation of motion and cannot be inferred from equilibrium thermodynamic considerations alone.

The transition from a discontinuous to a continuous transformation mode with decreasing temperature is also confirmed in this alloy. The annealing temperature $T^{*}_R=0.05$ used for Fig.\;\ref{fig6:kinetic_ordering_case_B_2} is low enough to place the kinetic pathway well within the continuous transformation range.

\subsubsection{\label{chap6_sec:level3_1_3}Alloy-3 (BCC$_{s.s.}$ $\Rightarrow$ B2)}
While FCC$_{s.s.}$ $\Rightarrow$ L1$_0$ ordering can be a 1st-order transition at relatively high temperatures, BCC$_{s.s.}$ $\Rightarrow$ B2 ordering is 2nd-order \cite{de1979configurational}. The difference can also be seen from the topology in the calculated phase diagrams (Fig.\;\ref{fig6:phase_diagram_ordering_three_cases}). While there is a two-phase region of a solid-solution and an ordered phase in Alloy-1 and Alloy-2, there are only single phase regions in Alloy-3. Alloy-3 is used to explore the kinetic difference between the 1st-order and 2nd-order phase transformations as well as the difference between the 2nd-order transition seen in the L1$_0$ ordering (i.e., spinodal ordering). 

\begin{figure}
\begin{center}
\includegraphics[scale=0.17]{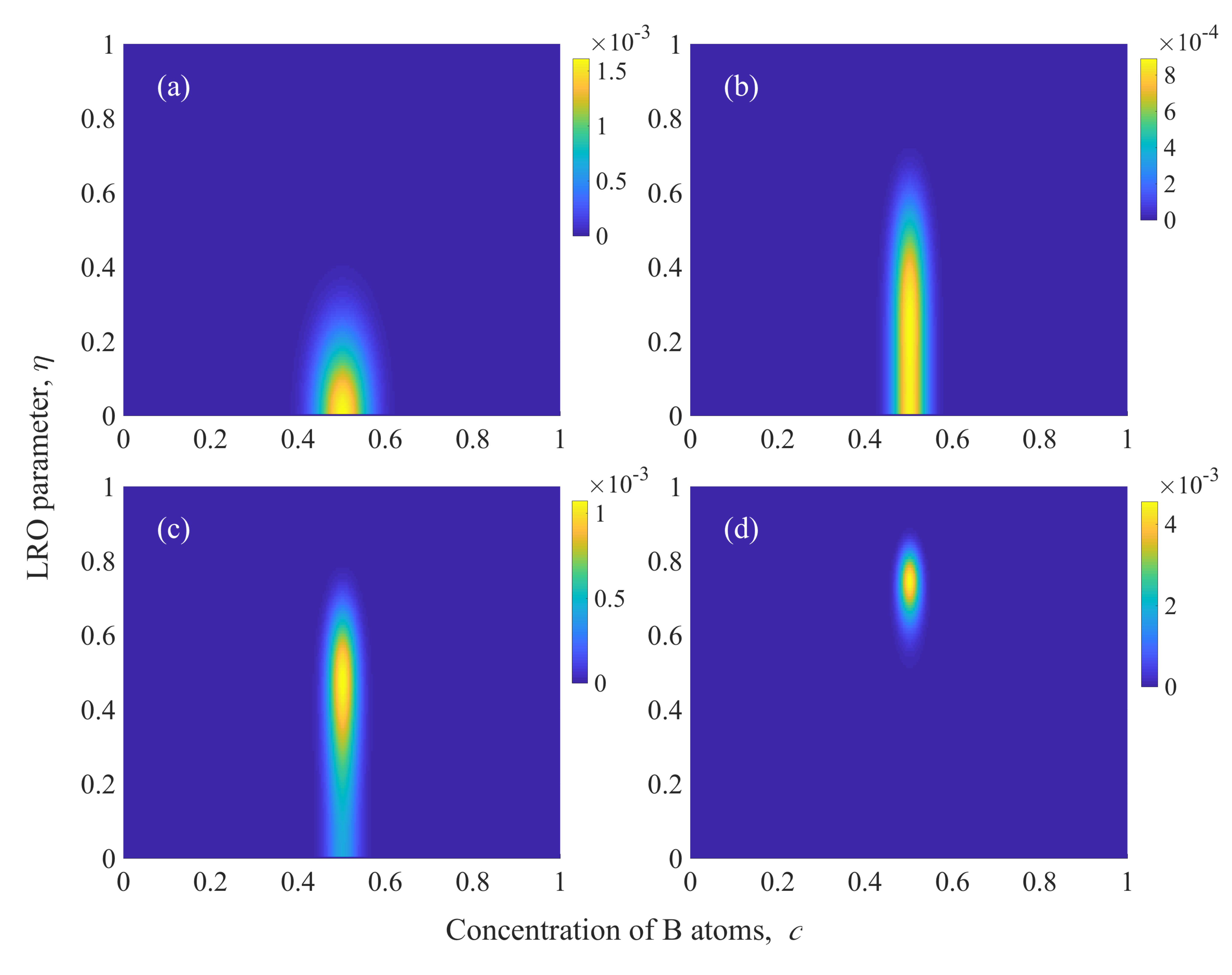}
\caption{\label{fig6:kinetic_ordering_case_B_3} The calculated kinetic ordering process in a A--50.0\;at.\%\;B alloy (Alloy-3 in Sec.\;\ref{chap6_sec:level3_1}) at $T^{*}_R=0.15$ using $N=10^4$ and $N_0=100$. The normalized time, $t^*$, of each snapshot is (a) $0.0000$, (b) $0.0080$, (c) $0.0100$, and (d) $0.0200$, respectively. }
\end{center}
\end{figure}

The calculated kinetic ordering process in a A--50.0\;at.\%\;B alloy system at $T^{*}_R=0.15$ is shown in Fig.\;\ref{fig6:kinetic_ordering_case_B_3}. Unlike the behavior during L1$_0$ ordering (Figs.\;\ref{fig6:kinetic_ordering_case_B_1_1}, \ref{fig6:kinetic_ordering_case_B_1_2}, and \ref{fig6:kinetic_ordering_case_B_2}), there is just one phase (or peak in probability space) during the phase transformation, and the single phase (or peak) moves from a disordered phase region into an ordered phase region continuously. This indicates that although both L1$_0$ ordering and B2 ordering at low annealing temperatures are 2nd-order and continuous, the kinetic behaviors are quite different.  

\subsection{\label{chap6_sec:level3_2}Concurrent phase separation and ordering}
\begin{figure}
\begin{center}
\includegraphics[scale=0.38]{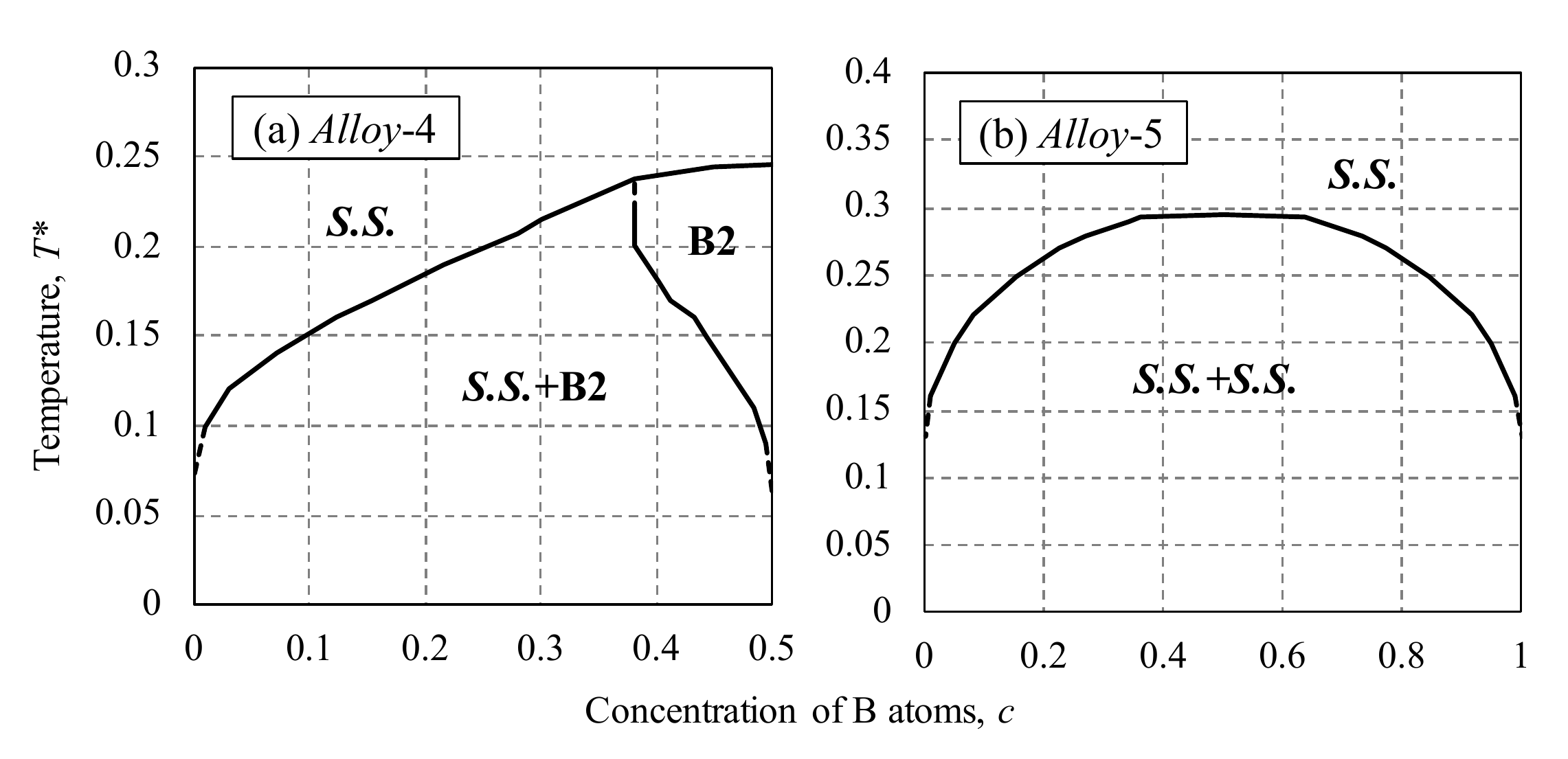}
\caption{\label{fig6:phase_diagram_phase_separation_ordering} The calculated phase diagrams for the two model alloy systems in Sec.\;\ref{chap6_sec:level3_2} using the SEAQT model. Alloy-4 has a two-phase region (BCC$_{s.s.}$\;$+$\;B2) at low temperatures, and Alloy-5 has a two-phase region (BCC$_{s.s.}$\;$+$\;BCC$_{s.s.}$) below the solvus line, which is a typical phase diagram seen in an alloy system which prefers just phase separation (see Part\;I \cite{yamada2018kineticpartI}. Here the temperatures are normalized as $T^*=k_BT/V(\bm{\mathrm{k}}_0)$, and the estimated lines are shown as broken lines. }
\end{center}
\end{figure}

In Section\;\ref{chap6_sec:level3_1}, the alloy systems that prefer just ordering are considered. In this section, alloy systems that have a tendency for both phase separation and ordering are explored by choosing positive $\omega$ in Eq.\;(\ref{eq6:total_energy_static_conc_2}). The values of $\omega$ for two more hypothetical alloy systems involving B2 ordering are shown in Table\;\ref{table6:different_cases_phase_separation_ordering}. No tetragonal distortions can arise from B2 ordering, so $\alpha=0$. The corresponding phase diagrams for these parameters calculated with the SEAQT theoretical framework are shown in Fig.\;\ref{fig6:phase_diagram_phase_separation_ordering}. For the kinetic calculations in this section, the normalized initial temperature is set as $T^{*}_0=0.5\; (T^{*}_0=k_B T/V(\bm{\mathrm{k}}_0))$ and some fluctuation is included in the initial states using $N_0=10^2$.
\begin{table}
\begin{center}
\caption{\label{table6:different_cases_phase_separation_ordering} The values of $\omega$ and $\alpha$ in the two model alloy systems for the phase separation and ordering calculations in Sec.\;\ref{chap6_sec:level3_2}. Both alloy systems involve B2 ordering on a BCC lattice. }
\begin{tabular}{ c  c  c } 
$\quad$ & $\quad$  & $\quad$  \\   \hline \hline
$\quad\quad\quad$  & \quad $\omega$ \quad \quad & \quad $\alpha$ \quad \quad \\ \hline
\quad Alloy-4 (B2 on BCC) \quad & 0.5 & 0.00 \\
\quad Alloy-5 (B2 on BCC) \quad & 1.2 & 0.00 \\ \hline \hline
\end{tabular}
\end{center}
\end{table}

\subsubsection{\label{chap6_sec:level3_2_1}Alloy-4 (BCC$_{s.s.}$ $\Rightarrow$ B2\;+\;BCC$_{s.s.}$) }
\begin{figure}
\begin{center}
\includegraphics[scale=0.17]{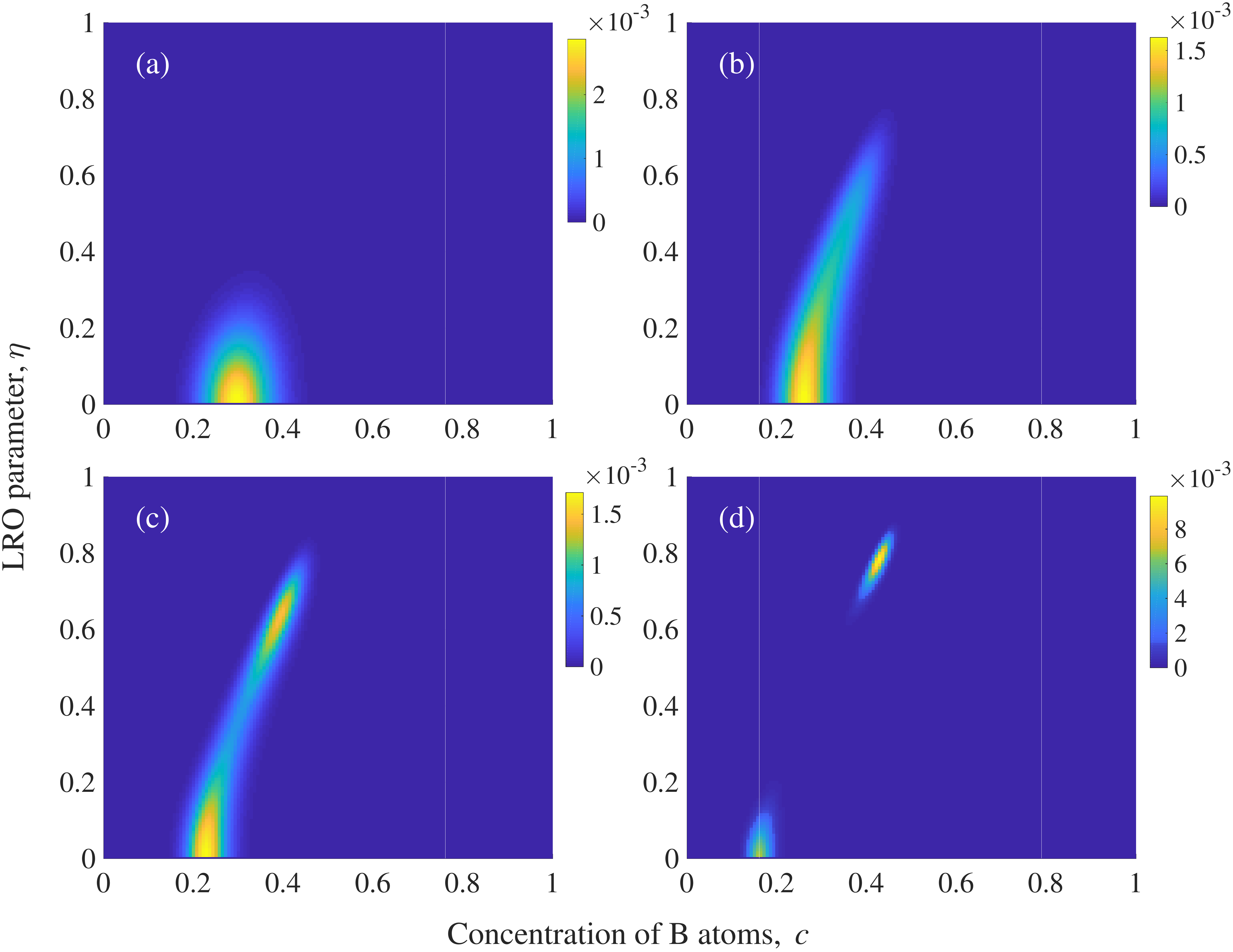}
\caption{\label{fig6:kinetic_phase_separation_ordering_C_1} The calculated kinetic evolution process in a A--30.0\;at.\%\;B alloy (Alloy-4 in Sec.\;\ref{chap6_sec:level3_2}) at $T^{*}_R=0.15$ using $N=10^4$ and $N_0=100$. The normalized time, $t^*$, of each snapshot is (a) $0.000$, (b) $0.017$, (c) $0.024$, and (d) $0.060$, respectively. }
\end{center}
\end{figure}

In this alloy, there is a two-phase region of the disordered BCC solid-solution and the B2 ordered phase in the calculated phase diagram (Fig.\;\ref{fig6:phase_diagram_phase_separation_ordering}\;(a)).  The calculated kinetic pathway in a A--30.0\;at.\%\;B alloy at $T^{*}_R=0.15$ is shown in Fig.\;\ref{fig6:kinetic_phase_separation_ordering_C_1}. The initial disordered solid-solution decomposes continuously and simultaneously into two phases: an ordered phase and an A-rich solid-solution. This behavior is described as concurrent ordering and phase separation and has been reported in the Fe--Be system \cite{ino1978pairwise}, whose experimentally determined phase diagram has similarities with the one calculated here (Fig.\;\ref{fig6:phase_diagram_phase_separation_ordering}\;(a)). Thus, the predicted kinetic path is qualitatively consistent with the reported experiments.

\subsubsection{\label{chap6_sec:level3_2_2} Alloy-5 (BCC$_{s.s.}$ $\Rightarrow$ BCC$_{s.s.}$\;+\;BCC$_{s.s.}$)}
\begin{figure}
\begin{center}
\includegraphics[scale=0.17]{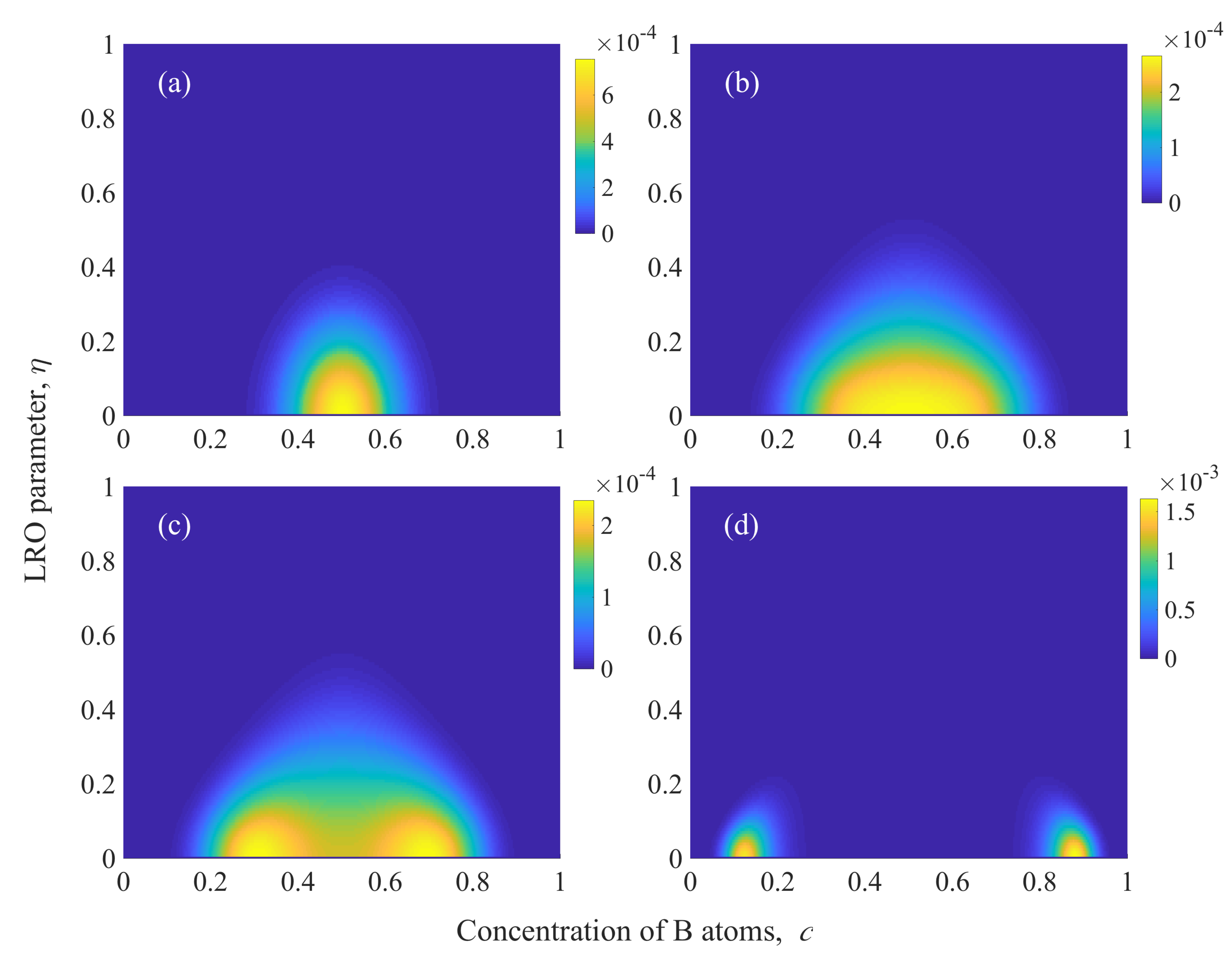}
\caption{\label{fig6:kinetic_phase_separation_ordering_C_2} The calculated kinetic evolution process in a A--50.0\;at.\%\;B alloy (Alloy-5 in Sec.\;\ref{chap6_sec:level3_2}) at $T^{*}_R=0.15$ using $N=10^4$ and $N_0=100$. The normalized time, $t^*$, of each snapshot is (a) $0.0000$, (b) $0.0040$, (c) $0.0048$, and (d) $0.0100$, respectively. }
\end{center}
\end{figure}

The calculated phase diagram shown in Fig.\;\ref{fig6:phase_diagram_phase_separation_ordering}\;(b) in this alloy system suggests a simple phase separation process of a solid-solution into two different solid-solutions with different compositions at low temperatures. However, the calculated kinetic behavior in this alloy system in a A--50.0\;at.\%\;B composition at $T^{*}_R=0.15$ (see Fig.\;\ref{fig6:kinetic_phase_separation_ordering_C_2}) demonstrates that ordering during the decomposition process can take place before the system ultimately reaches the final equilibrium state of two solid-solutions. The occupation probabilities for non-zero order parameters eventually disappear as the transformation proceeds. 

Ordered structures have been observed during the spinodal decomposition process in the Cu--10\;at.\%\;Co alloy \cite{busch1996high}, whose local average compositions are 50\;at.\%\;Co and 33\;at.\%\;Co. Although the lattice of 50\;at.\%\;Co has not been determined, it would suggest the importance of including longer-range interaction energies even in simple alloy systems with a positive mixing energy.

\section{\label{chap6_sec:level4}Conclusions}
The SEAQT model with a pseudo-eigenstructure based on the SCW method is applied to phase separation and ordering processes in a solid-solution in various binary model alloy systems, and the kinetic pathways are explored. The assumed model alloys are divided into two groups depending on  expected phase transformation behavior: ordering or both phase separation and ordering. While continuous and discontinuous ordering phenomena take place in the former group, concurrent phase separation and ordering are readily obtained in the latter.  

In the ordering calculations, while the B2 ordering shows only a continuous transformation mode, L1$_0$ ordering can take place both continuously and discontinuously depending upon the annealing temperature. Although both B2 and L1$_0$ ordering show continuous transformations, it turns out that their behaviors are quite different: there is only a single phase during B2 ordering, whereas two distinct phases evolve during L1$_0$ ordering. In addition, when elastic strain energy from a tetragonal distortion in the L1$_0$ ordered phase is small, it is found that a single L1$_0$ ordered phase decomposes into two different ordered phases at low temperatures. The calculated kinetic path of the decomposition process to the two different L1$_0$ ordered phases follows a ``solid-solution $\Rightarrow$ ordered phase $\Rightarrow$ two ordered phases" sequence.

In the phase separation and ordering calculations, concurrent phase separation and ordering is produced in a model alloy system, whose calculated phase diagram has similarities with the experimentally determined phase diagram in Fe--Be alloy system. Furthermore, an ordering behavior during the phase separation process is observed even though a simple phase separation process from a single solid-solution to two different solid-solutions is expected from the calculated phase diagram.   

Finally, the SEAQT framework has some distinct advantages for modeling kinetic behavior during alloy decomposition. It can describe kinetic paths from an initial non-equilibrium state to stable equilibrium without relying on a stochastic approach or a local/near equilibrium assumption, and alloy composition is automatically conserved during the kinetic calculations and kinetic paths involving two phases can be calculated in a single theoretical framework. 

\section*{ACKNOWLEDGEMENTS}
We acknowledge the National Science Foundation (NSF) for support through Grant DMR-1506936. \\

\begin{appendices}

\section*{\label{chap6_sec:level5}Appendix}

\section{\label{chap6_sec:level5_1}Estimation of phase diagrams with the SEAQT framework}
Phase diagrams are usually determined by calculating the free energies of candidate phases and searching for the phase with the lowest free-energy or searching for phases sharing the lowest common tangent to the molar free-energies. In this appendix, an approach to determine phase diagrams with the SEAQT framework without using free-energies is described. 

The occupation probabilities at stable equilibrium at $T^{\mathrm{se}}$ for a given pseudo-eigenstructure can be calculated from the (semi-) \cite{lesar2013introduction} grand canonical distribution:
\begin{equation}
p^{\mathrm{se}}_k=\frac{g_k e^{-\beta^{\mathrm{se}} ( E_k +\mu_A N_{A,k} +\mu_B N_{B,k} ) }}{\Xi} \; ,  \label{eq6:grand_canonical_distribution}
\end{equation}
where $\beta^{se}=1/k_BT^{\mathrm{se}}$, $\mu_A$ and $\mu_B$ are, respectively, the chemical potentials of A atoms and B atoms, and $\Xi$ is the grand partition function given by
\begin{equation}
\Xi \equiv \sum\limits_{k'} g_{k'} e^{-\beta^{\mathrm{se}} ( E_{k'} +\mu_A N_{A,k'} +\mu_B N_{B,k'} )}  \; . \label{eq6:grand_partition_function}
\end{equation}
The chemical potentials are adjusted to reach a target alloy composition. The stable equilibrium configuration can be found by searching a peak(s) in the probability distribution. For example, when peaks in the probability distributions appear in two regions of configuration space, two phases are present, and their concentrations are given by an average of the B concentrations of each peak region. The concentration of each phase is determined at each temperature considering a series of alloy compositions. Following statistical mechanical calculations for a solid phase that are quite large in size (a homogeneous system), a large system size, $N^{\mathrm{se}}=10^{10}$, is used here for the calculations. 

\end{appendices}

\bibliographystyle{ieeetr}
\bibliography{ref}

\end{document}